\documentclass[lengthestimate,tighten,amssymb,prb,aps,twocolumn,floatfix,tightenlines,superscriptaddress]{revtex4-1}
\usepackage{graphicx,subfigure,color,epstopdf}
\usepackage[T1]{fontenc}
\usepackage[latin9]{inputenc}
\usepackage{longtable}
\usepackage{graphicx}
\usepackage{color}



\begin{document}
\title{Quantum Phase Estimation Algorithm for Finding Polynomial Roots}
\author{T. Tansuwannont}
\affiliation{Department of Physics, Faculty of Science, Chulalongkorn University \\
254 Phayathai Road, Pathumwan, Bangkok, 10330, Thailand}\affiliation{Collaborative Research Unit on Quantum Information \& Department of Physics, Faculty of Science, Mahidol University, Rama VI Road, Ratchathewi, Bangkok, 10400, Thailand}

\author{S. Limkumnerd}
\affiliation{Department of Physics, Faculty of Science, Chulalongkorn University \\
	254 Phayathai Road, Pathumwan, Bangkok, 10330, Thailand}\affiliation{
Research Center in Thin Film Physics, Thailand Center of Excellence in Physics, CHE, 328 Si Ayutthaya Rd., Bangkok 10400, Thailand}

\author{S. Suwanna}
\affiliation{Collaborative Research Unit on Quantum Information \& Department of Physics, Faculty of Science, Mahidol University, Rama VI Road, Ratchathewi, Bangkok, 10400, Thailand}

\author{P. Kalasuwan}
\email{Corresponding author's email: pruet.k@psu.ac.th}\affiliation{Department of Physics, Faculty of Science, Prince of Songkla University, Hat-Yai, Songkhla, 90112, Thailand}

\date{\today}

\begin{abstract}
Quantum algorithm is an algorithm for solving mathematical problems using quantum systems
encoded as information, which is found to outperform classical algorithms in some specific cases. The objective of this
study is to develop a quantum algorithm for finding the roots of nth degree polynomials where
n is any positive integer. In classical algorithm, the resources required for solving this problem
increase drastically when n increases and it would be impossible to practically solve the problem
when n is large. It was found that any polynomial can be rearranged into a corresponding
companion matrix, whose eigenvalues are roots of the polynomial. This leads to a possibility
to perform a quantum algorithm where the number of computational resources increase as a
polynomial of n. In this study, we construct a quantum circuit representing the companion
matrix and use eigenvalue estimation technique to find roots of polynomial.

\end{abstract}

\maketitle

\section{Introduction}

Roots finding is a centuries-old problem that has continued to attract considerable research interests and efforts due to its relevance in many fields of mathematics and physics involving geometry, number theory, probability and combinatorics. It is well known that for a polynomial of degree 4 or or less, there exists a formula or procedure to solve for its roots exactly \cite{Weisstein-crc-2009}. However, such a task is impossible for a polynomial of degree 5 or greater \cite{Jacobson-dover-2009}. Many root-finding algorithms have been devised for obtain approximated roots of a polynomial of arbitrary degree \cite{McNamee-elsvr-2007,Mekwi01,pan-siam-39-187}.  

The plausibility of a quantum computer---a new type of computation, which embraces quantum mechanics into its information, algorithms, and output measurements---has entailed quantum algorithms. A simple enhancement by rather non-intuitive mathematics of quantum mechanics like superposition, the uncertainty principle, and entanglement, brings quantum algorithms forth to a new level of computation unreachable before by conventional computers with classical algorithms. For example, the Shor's algorithm, an algorithm to factorize a large integer into a product of primes, is proven in principles to overcome the classical-algorithm limit in terms of speed\cite{sh-conf-94-124}. A breakthrough in quantum simulation is expected to bring eminent impact into science and technology\cite{ll-sci-273-1073,fe-ijtp-82-467}. 
Recent progress on the actual quantum devices, such as a successful small-molecule simulation \cite{as-sci-309-1704,ma-natphys-7-399,brn-prl-97-050504,neg-pra-71-032344} or probing the statistics of quantum systems\cite{mtt-srep-3-1539,ovi-jsm-2015-1-P01004,crsp-pra-91-013811}, have yielded high promises and attracted immense interests. The advancement of computation and simulation in the aforementioned examples owes largely to a common underlying method called \emph{phase estimation algorithm} (PEA)\cite{kitaev96}. Still, PEA can be improved, especially at the fundamental algorithm of finding roots of a polynomial.


However, PEA is only applicable to unitary operators which are not always the case for some quantum algorithms; for instance, the phase measurement under the circumstance where decoherence is present in the process\cite{kac-nphot-4-357360}. In a measurement process of the quantum algorithm, the non-unitary matrices also play key roles as projective operators. In order to modify existing PEA to be suitable for eigenvalue problems comprehensively, a programmable circuit, and measurement of the control and ancillary qubits are recently exploited to tailor-made any arbitrary matrix\cite{dask-qip-13-333}. The great advantage of this proposed scheme is that any matrix can be constructed and the control gate of the respective matrix can be realized, paving  ways to build a quantum computer which can calculate eigenvalue of any matrix. However, some drawbacks exist as the algorithm itself may not be efficient for complicated matrices, which quantum complexity arises following the increasing number of non-zero matrix elements\cite{dask-jchemphy-137-234112}. Further investigation on the algorithm in terms of appropriate complexity is still needed. 

Our main aim in this paper is to propose a modified quantum phase estimation algorithm for finding polynomial roots, where we present a benchmark implementation of quantum non-unitary eigenvalue calculation scheme for polynomials. This specific task represents the least complex eigenvalue, which the algorithm can be fruitful without too much concern over the complexity. 

This article is organized as follows. The remaining subsections of this section will cover key concepts and ideas about the phase estimate algorithm, and the iterated phase estimate algorithm (IPEA) for unitary operators, as well as the quantum algorithm to find complex eigenvalues of a general matrix. In Section II, we present our modified PEA and IPEA, together with the companion matrix approach, and more importantly, the circuit design to estimate roots of a polynomial of degree $n$. There we focus our presentation of the circuit operation and outputs, leaving the discussion and complexity analysis in Section III. Finally, the conclusions are summarized in Section IV.

\subsection{PEA and IPEA for Eigenvalue Problems of Unitary Operators} 
In the original version of PEA, a phase $\varphi$ arising after a unitary evolution $U$ with eigenvalue $\exp(2\pi\varphi i)$ is operated on its basis. Because a quantum evolution can be interpreted by a phase factor $U=\exp(-i\mathbf{H}t/\hbar)$, where $\mathbf{H}$ is a Hamiltonian of a finite system, the phase as a result of the phase estimation algorithm is indeed the eigenvalue of the Hamiltonian. PEA has also been introduced as a potential quantum tool to effectively solve various eigenvalue problems involving unitary operators\cite{da-prl-83-5162,zhxq-nphot-7-223228,wng-pra-82-062303,brz-scr-4-6115}. The unitary operators play a central role in all of the quantum algorithms, as they are required for universal quantum computations\cite{ba-pra-52-3457}. 

\begin{figure}[htbp]
    \centering
    \includegraphics[width=0.5\textwidth]{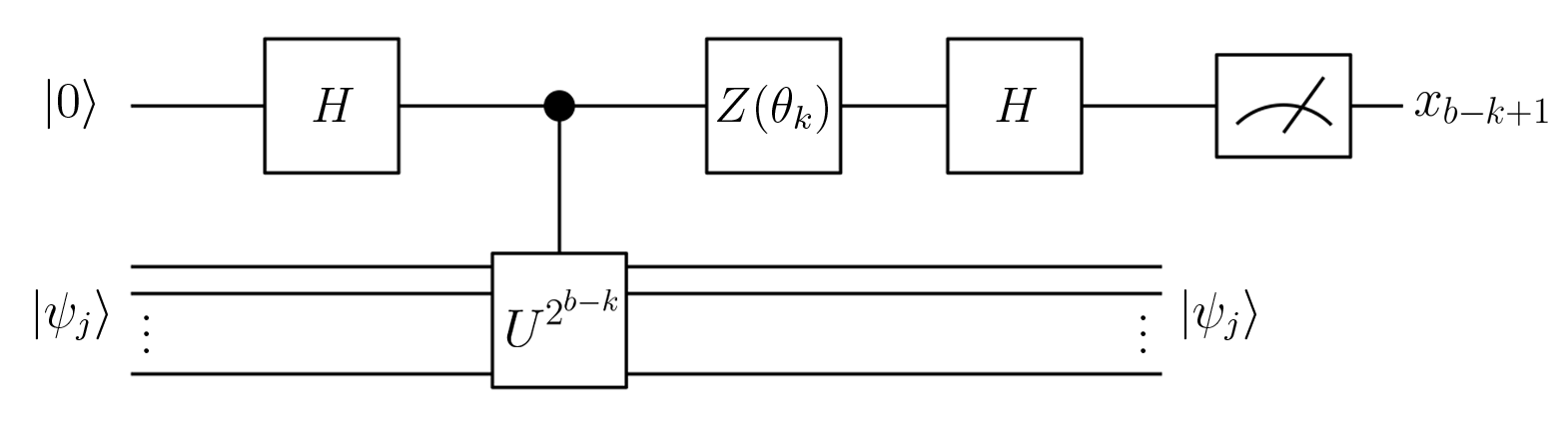}
    \caption{ A circuit for the $k^{\mathrm{th}}$ iteration of the IPEA where $\theta_k$ is the feedback from prior iterations.}
    \label{fig:1}
\end{figure}

In order to estimate the value of a phase parameter $\omega_j$ up to the $b$ bit-precision using PEA,
$b$ ancillary qubits in control register are required. In practice, however, the number of qubits which can
be implemented is very limited. Iterative Phase Estimation Algorithm (IPEA) is an algorithm
improved from the original PEA with an aim to estimate $\omega_j$ up to $b^{\text{th}}$ digit while using only one
ancillary qubit together with $b$ iterations as a result of scalable inverse quantum Fourier transform in a semi-classical manner\cite{gr-prl-76-3228,ch-sci-308-997}. In order to explain the algorithm as illustrated in Fig. \ref{fig:1}, we first assume that the phase parameter $\omega_j$ has a binary expansion no more than $b$ digits (written as $\omega_j = 0.x_1x_2x_3\ldots x_b000\ldots$). Initially, all of the ancillary qubits are prepared in state $|0\rangle$ and the target register is prepared in the eigenstate $|\psi_j\rangle$ of unitary operator $U$. A Hadamard gate is applied to the control register in order to prepare state $\left(\frac{|0\rangle+|1\rangle}{\sqrt{2}}\right)$. In the first iteration $(k=1)$, a c-$U^{2^{b-1}}$ and 
\begin{equation}
                      Z\left(\theta_k\right)=\left[\begin{array}{cc} 1 & 0 \\
                           0 & \text{e}^{-i\theta_k}\end{array}\right], 
                  \label{eq:ztheta}
                 \end{equation}
  where $\theta_1=0$ are applied. After that, the second Hadamard gate is applied on the control qubit and its state is measured in the computational basis $\{|0\rangle,|1\rangle\}$. This results in state
  \begin{equation}
\frac{1}{2}\big[(1+\text{e}^{2\pi i(0.x_b)})|0\rangle+(1-\text{e}^{2\pi i(0.x_b)})|1\rangle\big],
\label{eq:ipeaprob}
\end{equation}  
whose measurement gives either $0$ or $1$, and is determined by the majority probability between $|0\rangle$ and $|1\rangle$. This measurement result consequently dictates the value of $x_b$. The next iteration is performed with the c-$U^{2^{b-k}}$ and $Z\left(\theta_k\right)$, where $\theta_k=2\pi\left(0.0x_{b-k+2}x_{b-k+3}\ldots x_b\right)$ is calculated by the feed-forwarded measurement result of the prior iterations up to $x_{b-k+1}$. The algorithm is finished when the digit $x_1$ is obtained.

The original IPEA has been used to determine only the phase parameter $\omega_j$ of the eigenvalue $\lambda_j = \text{e}^{2\pi i \omega_j}$ of a unitary matrix $U$. In general, however, an eigenvalue of a non-unitary operator can be written as 
$\lambda_j = |\lambda_j|\text{e}^{2\pi i \omega_j}$. The standard IPEA therefore cannot be applied without the knowledge of  modulus $|\lambda_j|$.


\subsection{Quantum Algorithm for Finding Complex Eigenvalues of General Matrices}
Recently,  Daskin \textit{et al.} have introduced their technique to find the complex eigenvalues of general matrices\cite{dask-qip-13-333}. 
In order to employ the IPEA on the non-unitary operators, first of all, the non-unitary operator $O$ has to be controlled by a phase qubit and the c-$U^{2^{b-k}}$ in Fig. \ref{fig:1} is replaced by c-$O^{2^{b-k}}$. 

	
In the scheme proposed by Daskin \textit{et al.}, the decomposition of the control gate of the non-unitary operator $O$ of size $N=2^m$ uses the programmable circuit design, which requires $m+1$ ancillary qubits and $m$ main qubits\cite{dask-qip-13-333}. The operator $O$ generally has a eigenvalue of the form $|\lambda_j|\text{e}^{2\pi i \omega_j}$. In case that $\omega_j=0.x_1000\ldots$, an operation of c-$O$ followed by the Hadamard gate gives an output state
\begin{equation}
\left[\!\left(1+|\lambda_j|\text{e}^{2\pi i (0.x_1)}\!\right)|0\rangle_p\! + \!\left(1-|\lambda_j|\text{e}^{2\pi i (0.x_1)}\right)|1\rangle_p \right] \!|\psi_j\rangle_m,
\label{eq:findaskin}
\end{equation}
where $p$ and $m$ denote phase qubit and main qubits, respectively. As can be seen, the c-$O^{2^{b-k}}$ can be realized by the decomposition proposed by Daskin \textit{et al.} It is also possible to estimate its phase $\omega_j$ via the IPEA from the probability shown in (\ref{eq:findaskin}) 
in the same fashion as the phase estimation results determined by (\ref{eq:ipeaprob}).
However, the 
true novelty of the scheme is in the estimation of $|\lambda_j|$---taking the calculation 
to a complete eigenvalue estimation for any non-unitary matrix. Following the result shown in (\ref{eq:findaskin}), the value of $|\lambda_j|$ is related to the probability $P_0$ or $P_1$ of finding the phase qubit in states $|0\rangle$ or $|1\rangle$, respectively. Let $P=\max\{P_0,P_1\}$, so that we can estimate $|\lambda_j|$ as
\begin{equation}
|\lambda_j|=2N^2\sqrt{P}-1,
\end{equation}
where $N$ is the dimension of matrix. In practice, $|\lambda_j|$ is determined by the statistics of the measurement. We can also improve the accuracy of the estimation by using the statistics from other iterations. For the $k^\mathrm{th}$ iteration after which c-$O^{2^{b-k}}$ is operated followed by $Z(\theta_k)$ and the Hadamard gate, the relationship between $P^{(k)}$ and $|\lambda_j|$ becomes
\begin{equation}
|\lambda_j|^{2^{b-k}}=2N^2\sqrt{P^{(k)}}-1.
\end{equation}
Since we can estimate both $|\lambda_j|$ and $\omega_j$, the complex eigenvalue $\lambda_j$ can be determined.

\section{Quantum Algorithm for Finding Polynomial Roots}
\subsection{Companion Matrix Approach}
As the aim of this study is to find roots of a generic polynomial of degree $n$, we can formulate this problem as the eigenvalue problem of a non-unitary operator. First of all, consider 
\begin{equation}
p(x) = x^n+a_{n-1}x^{n-1}+\cdots+a_1x+a_0;
 \label{eq:plynoml}
\end{equation}
it can be factorized into the form 
\begin{equation}
p(x) =(x-z_1)(x-z_2)\cdots(x-z_n), 
 \label{eq:facply1}
\end{equation}
where $z_1, z_2,\ldots, z_n \in\mathbb{C}$ are the roots of $p(x)$. From general linear algebra\cite{buchberger1985}, the roots of polynomial $p(x)$ are eigenvalues of its companion matrix defined as
\begin{equation}
	 C_p=\left[\begin{array}{ccccc} 0 & 1 & 0 & & \\
     & 0 & 1 & 0 & \\ & & \ddots & \ddots & \\ & & & 0 & 1\\ -a_0 & -a_1 & \cdots & -a_{n-2} & -a_{n-1}\end{array}\right]  
     \label{eq:cp}
\end{equation}
with respect to the basis $\left\{ 1,x,x^2,\ldots,x^{n-1}\right\}$.
                 

Daskin's algorithm requires that the absolute value of every coefficient $a_i$ must be less than or equal to 1, since rotation gates are used to simulate these coefficients. Therefore, we introduce a scaling method to meet this requirement. Let $a_{\mathrm{max}}$ denote the greatest absolute value of $a_0,a_1,\ldots,a_n$. We choose a basis of circuit in the $x$-mode or $(1/x)$-mode depending on whether $|a_n|$ or $|a_0|$ is greater to maximize the success probability of the circuit scheme.

In case $|a_n|>|a_0|$, the $x$-mode will be chosen, so the polynomial $p(x)$ can be equivalently expressed in the form
\begin{equation}
\frac{a_n}{a_{\mathrm{max}}}x^n+\frac{a_{n-1}}{a_{\mathrm{max}}}x^{n-1}+\cdots+\frac{a_1}{a_{\mathrm{max}}}x+\frac{a_0}{a_{\mathrm{max}}}=0.
\end{equation}
Let $\mu=\frac{a_{\mathrm{max}}}{a_n}$ be a scaling factor. Then the corresponding eigenvalue equation is written as        
\begin{eqnarray}
&&\hspace*{-0.5cm}\left[\begin{array}{ccccc} 
0 & \frac{a_n}{a_{\mathrm{max}}} & 0 & & \vspace*{1mm}\\
& 0 & \frac{a_n}{a_{\mathrm{max}}} & 0 & \\
&   &\ddots & \ddots & \\
&   &   & 0 & \frac{a_n}{a_{\mathrm{max}}} \vspace*{1mm}\\
-\frac{a_0}{a_{\mathrm{max}}} & -\frac{a_1}{a_{\mathrm{max}}} & \cdots & -\frac{a_{n-2}}{a_{\mathrm{max}}} & -\frac{a_{n-1}}{a_{\mathrm{max}}}
\end{array}\right] \left[\begin{array}{c}
1 \\ x \\ \vdots \vspace*{2mm}\\ x^{n-2} \vspace*{1mm}\\ x^{n-1} \vspace*{1mm}
\end{array}\right] \nonumber\\
&=&\left[\begin{array}{ccccc} 
0 & \frac{1}{\mu} & 0 & & \vspace*{1mm}\\
  & 0 & \frac{1}{\mu} & 0 & \\
  &   &\ddots & \ddots & \\
  &   &   & 0 & \frac{1}{\mu} \vspace*{1mm}\\
-{a'_0} & -{a'_1} & \cdots & -{a'_{n-2}} & -{a'_{n-1}}
\end{array}\right] \left[\begin{array}{c}
1 \\ x \\ \vdots \vspace*{2mm}\\ x^{n-2} \vspace*{1mm}\\ x^{n-1} \vspace*{1mm}
\end{array}\right] \nonumber\\
&=&\frac{x}{\mu}
\left[\begin{array}{c}
1 \\ x \\ \vdots \vspace*{2mm}\\ x^{n-2} \vspace*{1mm}\\ x^{n-1} \vspace*{1mm}
\end{array}\right].
\label{eq:scaling1}
 \end{eqnarray}  
  where $a'_i=\frac{a_i}{a_\mathrm{max}}$. The eigenvalue of modified companion matrix is $x/\mu$.
   
On the other hand, if $|a_0|>|a_n|$, the $(1/x)$-mode will be used. Dividing the polynomial $p(x)$ by $a_{\mathrm{max}}x^n$ leads to
\begin{equation}
\frac{a_0}{a_{\mathrm{max}}}\!\left(\frac{1}{x}\right)^n + \frac{a_1}{a_{\mathrm{max}}}\!\left(\frac{1}{x}\right)^{n-1}\!\!+ \cdots+ \frac{a_{n-1}}{a_{\mathrm{max}}}\left(\frac{1}{x}\right)+\frac{a_n}{a_{\mathrm{max}}}=0.
\end{equation}
In this case, a scaling factor is $\mu=\frac{a_{\mathrm{max}}}{a_0}$, and the corresponding eigenvalue equation is in the form 
\begin{eqnarray}
\hspace*{-0.5cm}&&\left[\begin{array}{ccccc} 
0 & \frac{a_0}{a_{\mathrm{max}}} & 0 & & \vspace*{1mm} \\
& 0 & \frac{a_0}{a_{\mathrm{max}}} & 0 & \\
&   &\ddots & \ddots & \\
&   &   & 0 & \frac{a_0}{a_{\mathrm{max}}} \vspace*{1mm}\\
-\frac{a_n}{a_{\mathrm{max}}} & -\frac{a_{n-1}}{a_{\mathrm{max}}} & \cdots & -\frac{a_{2}}{a_{\mathrm{max}}} & -\frac{a_{1}}{a_{\mathrm{max}}}
\end{array}\right] \left[\begin{array}{c}
1 \\ 1/x \\ \vdots \vspace*{1mm}\\ 1/x^{n-2} \vspace*{1mm}\\ 1/x^{n-1} \vspace*{1mm}
\end{array}\right] \nonumber\\
&=&\left[\begin{array}{ccccc} 
0 & \frac{1}{\mu} & 0 & & 1 \vspace*{1mm}\\ 
  & 0 & \frac{1}{\mu} & 0 & \\
  &   &\ddots & \ddots & \\
  &   &   & 0 & \frac{1}{\mu} \vspace*{1mm}\\
-{a'_0} & -{a'_1} & \cdots & -{a'_{n-2}} & -{a'_{n-1}}
\end{array}\right] \left[\begin{array}{c}
1 \\ 1/x \\ \vdots \vspace*{1mm}\\ 1/x^{n-2} \vspace*{1mm}\\ 1/x^{n-1} \vspace*{1mm}
\end{array}\right] \nonumber\\
&=&\frac{1}{\mu x}
\left[\begin{array}{c}
1 \\ 1/x \\ \vdots \vspace*{1mm}\\ 1/x^{n-2} \vspace*{1mm}\\ 1/x^{n-1} \vspace*{1mm}
\end{array}\right],
\label{eq:scaling2}
 \end{eqnarray}
 where $a'_i=\frac{a_{n-i}}{a_\mathrm{max}}$ in this case. Note that the eigenvalue of the modified companion matrix is $1/\mu x$.
 
However, the traditional companion matrix as described in (\ref{eq:cp}) has 1's in the upper diagonal entries but all of such entries of the modified companion matrices as shown in (\ref{eq:scaling1}) and (\ref{eq:scaling2}) have absolute values less than 1. To rectify this, we introduce a scaling gate $S_{m,\mu}$ which will be explained in details later; see Equation (\ref{eq:Sgate}).
 

\subsection{Quantum Circuit Design}
\begin{figure*}[t!]
\includegraphics[width=0.75\textwidth]{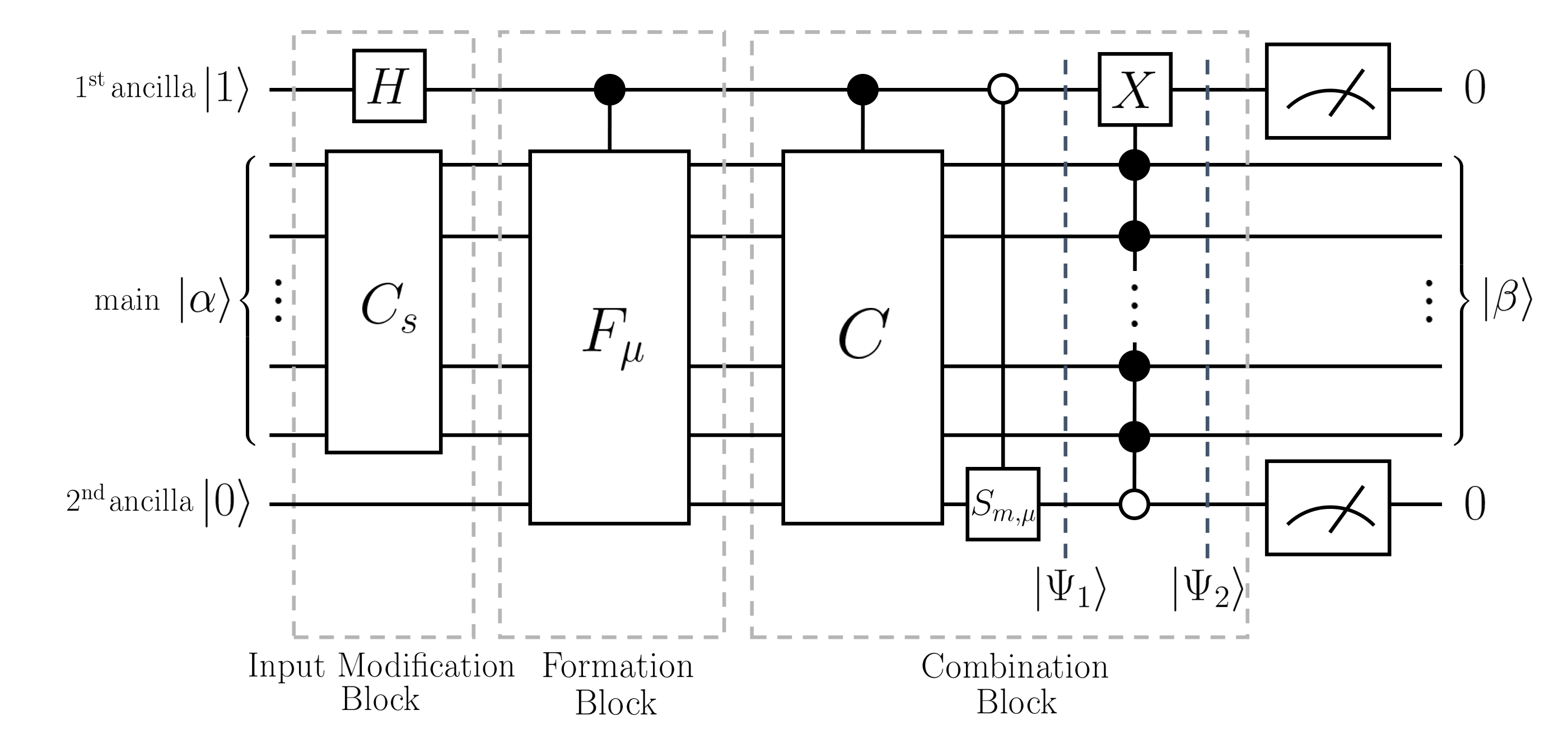}
\caption{Polynomial Representative Circuit (PRC) which is used to represent an operation of a companion matrix.}
\label{fig:2} \vspace{-0.4cm}
\end{figure*}

\begin{figure}[htbp]
    \centering
    \includegraphics[width=0.35\textwidth]{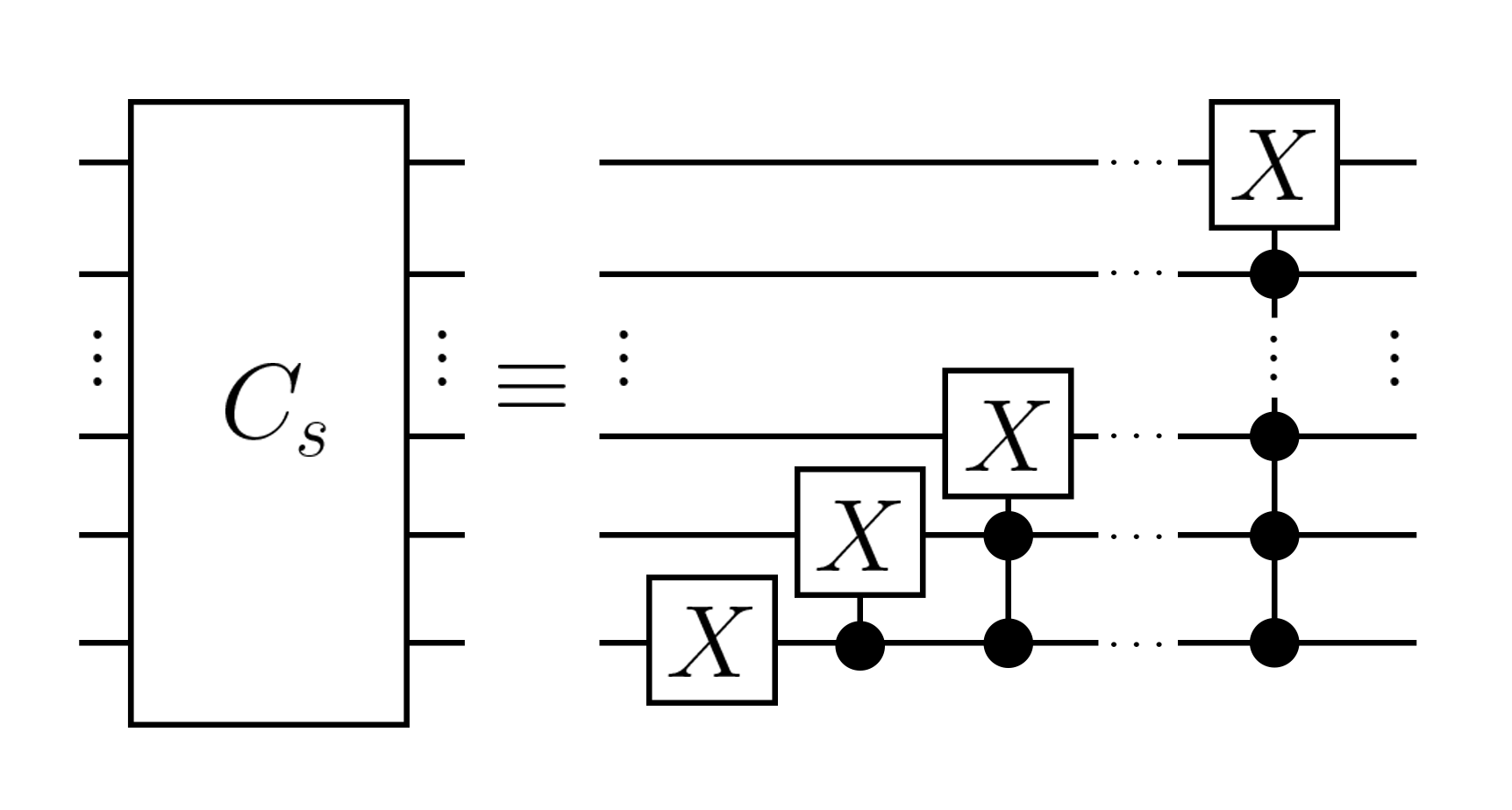}
    \caption{The cyclic-swap gate can be implemented by a sequence of Toffoli gates.}
    \label{fig:3}
\end{figure}

Our design of the respective algorithm relies on Polynomial Representative Circuit (PRC), a circuit to represent this modified companion matrix as illustrated in Fig. \ref{fig:2}. PRC requires $m$ main qubits and 2 ancillary qubits where $2^m=n$ is a degree of the polynomial. (Although it is inconvenient, the circuit is also applicable for $n \neq 2^m$ simply by shifting the degree of polynomial up to the nearest power of 2.)
First, let the main qubit be prepared in the initial state : 
\begin{eqnarray}
|\alpha\rangle =\left[\begin{array}{c}
\alpha_0 \\ \alpha_1 \\ \vdots \\ \alpha_{n-2} \\ \alpha_{n-1}
\end{array}\right],
\end{eqnarray} 
and we define $|\beta\rangle$ as a result of $C_p$ operating on $|\alpha\rangle$; i.e.
\begin{equation}
C_p|\alpha\rangle = |\beta\rangle.
\end{equation}
Multiplying $|\alpha\rangle$ by the modified companion matrix from (\ref{eq:scaling1}) or (\ref{eq:scaling2}) gives $|\beta\rangle$ in the form:
\begin{equation}
|\beta\rangle = \left[\begin{array}{c}
\alpha_1 \\ \alpha_2 \\ \vdots \\ \alpha_{n-1} \\ -\vec{a'}\cdot\vec{\alpha}
\end{array}\right],
\end{equation}
where $\vec{a'}\cdot\vec{\alpha}=a'_0\alpha_0+a'_1\alpha_1+\cdots+a'_{n-1}\alpha_{n-1}$. Similar to the circuit introduced by Daskin \textit{et al.}, our circuit consists of \textit{Input Modification Block, Formation Block}, and \textit{Combination Block}. The main ingredient of the Input Modification Block is a cyclic-swap gate $C_s$ applied on the main qubits. The matrix representation of the gate is 
\begin{equation}
 C_s=\left[\begin{array}{ccccc} 0 & 1 & 0 & \cdots & 0 \\
 0 & 0 & 1 & \cdots & 0\\ & & \ddots & \ddots & \\0 & 0 & \cdots & 0 & 1\\ 1 & 0 & \cdots & 0 & 0\end{array}\right], 
 \label{eq:cs}
 \end{equation}
 which can be implemented by the Toffoli gates as shown in Fig. \ref{fig:3}.

The aim of the operator $C_s$ is to generate the matrix element of the companion matrix from row $1$ to row $n-1$. Since the operation is underpinned by the presences of the sequences of Toffoli gates, the algorithm will be plagued by the huge complexity. the complexity of the algorithm is quite large, and yet still smaller than that of Daskin's scheme, as their matrix elements are generated by the formation block which incurs more complexity. 

\begin{figure}[htbp]
 	\centering
 	\includegraphics[width=0.45\textwidth]{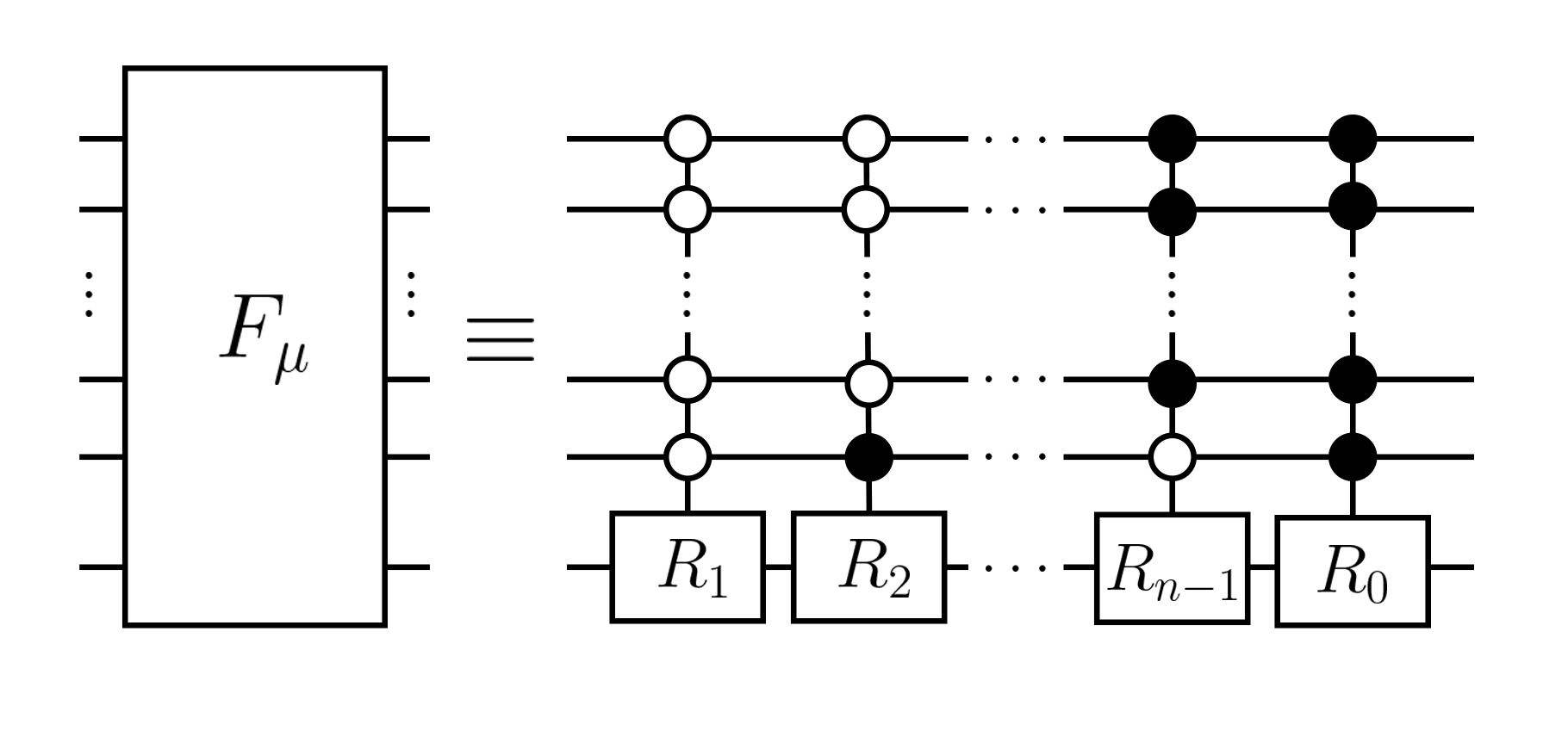}
 	\caption{The formation block can be simulated by a sequence of controlled-rotation gates.}
 	\label{fig:4}
 \end{figure}

The formation block in our version plays a role of the controlled gate of an operator $F_{\mu}$, which represents the components in the last row of the modified companion matrix $C_p$ as in (\ref{eq:scaling1}) or (\ref{eq:scaling2}). The rotation gate $R_i$ is represented by a matrix as follows:
\begin{eqnarray}
\label{eq:rotation}
R_{i}=
\left[\begin{array}{cc}
a'_i & \sqrt{1-{a'_i}^2} \\
-\sqrt{1-{a'_i}^2} & a'_i
\end{array}\right],
\end{eqnarray}
where $i=0,1,2,\ldots,n-1$.
Accordingly, the array of $R_i$ forms the block matrix $F_{\mu}$ as follows: 
\begin{equation}
                      F_{\mu}=\left[\begin{array}{cccc} R_1 & & & \\
                          & \ddots &  & \\ & & R_{n-1} & \\ & & & R_0 \end{array}\right].  
                  \label{eq:fff}
                 \end{equation}
The operation of $F_{\mu}$ can be simulated by a sequence of controlled-rotation gates as in Fig. \ref{fig:4}. The operation of $F_{\mu}$ will be performed on main qubits and the second ancillary qubit in case that the state of first ancillary qubit is $|1\rangle$.
 
Next step, in the combination block, we define the operator $C$ as follows:
\begin{equation}
C=\left(XH\right)^{\otimes m}\otimes I \nonumber \\
=\left[\begin{array}{ccccccc} \bullet & \bullet & \bullet & \bullet & \cdots & \bullet & \bullet \\
\bullet & \bullet & \bullet & \bullet & \cdots & \bullet & \bullet \\
\vdots & \vdots & \vdots & \vdots & \ddots & \vdots & \vdots \\
\bullet & \bullet & \bullet & \bullet & \cdots & \bullet & \bullet \\
1 & 0 & 1 & 0 & \cdots & 1 & 0 \\
\bullet & \bullet & \bullet & \bullet & \cdots & \bullet & \bullet
 \end{array}\right].
\end{equation}
where "$\bullet$" represents the elements we can neglected because they will be filter out by post-selection at the final stage of the algorithm. 

There are three sub-tasks to undertake in the combination block. First, a controlled-$C$ gate with the operator $C$ is operated on the main qubits conditioning to the state of the first ancillary qubit as $|1\rangle$ to create the $\vec{a'}\cdot\vec{\alpha}$ component. As a result, the operation of $C_s$, $F_\mu$, and $C$ on main qubits and second ancillary qubit conditioning to the first ancillary state $|1\rangle$ give the following output:
\begin{equation}
\frac{1}{\sqrt{2^m}}\left[\begin{array}{c}
\bullet \\ \bullet \\ \vdots \\ \bullet \\ \vec{a}\cdot\vec{\alpha} \\ \bullet \end{array}\right]
\end{equation}
with the probability amplitude $\frac{1}{\sqrt{2^m}}$. This amplitude is required to be balanced with the case where the state of the first ancillary qubit is $|0\rangle$. At this stage of the algorithm, the scaling gate $S_{m,\mu}$ is introduced to balance this probability and generate $1/\mu$ in the upper diagonal entries of the modified companion matrix. It is defined as
\begin{equation}
S_{m,\mu}=\frac{1}{\sqrt{2^m}\mu}\left[\begin{array}{cc} 1 & \sqrt{2^m\mu^2-1} \\                           -\sqrt{2^m\mu^2-1}& 1 \end{array}\right]. \label{eq:Sgate}
\end{equation}
This gate is, in fact, a rotation gate $R_y(\theta)$, where $\theta = 2\cos^{-1}(1/\sqrt{2^m}\mu)$. Scaling gate $S_{m,\mu}$ is to be operated on the second ancillary qubit in case that first ancillary qubit is in state $|0\rangle$. After the controlled-$S_{m,\mu}$ gate, the state is transformed into the following, 

\begin{eqnarray}
|\Psi_1\rangle  &=& \frac{1}{\sqrt{2}}\Big[|0\rangle\otimes \left(\left(I^{\otimes m} \otimes S_{m,\mu}\right)\left(C_s\otimes I\right)\left(|\alpha\rangle\otimes|0\rangle\right) \right) \nonumber \\ && -|1\rangle\otimes CF_{\mu}\left(C_s\otimes I\right)\left(|\alpha\rangle\otimes|0\rangle\right)\Big] \nonumber\\
&=& \frac{1}{\sqrt{2^{m+1}}\mu}\Big[|0\rangle\otimes\begin{array}{ccccccccc}[\alpha_1 & \bullet & \alpha_2 & \bullet & \cdots & \alpha_{n-1} & \bullet & \alpha_0 & \bullet]\end{array} \nonumber\\ && +|1\rangle\otimes\begin{array}{cccccc}[\bullet & \bullet & \cdots & \bullet & -\vec{a'}\cdot\vec{\alpha} & \bullet]\end{array}\Big]. \label{eq:afterscale}
\end{eqnarray}

However, referring to (\ref{eq:afterscale}), the final state is not exactly $|\beta\rangle$. The last task is just to swap between the coefficient $\alpha_0$ and $-\vec{a'}\cdot\vec{\alpha}$ in (\ref{eq:afterscale}) using the Toffoli gate in Fig. \ref{fig:2}, which results in
\begin{equation}
|\Psi_2\rangle=\frac{1}{\sqrt{2^{m+1}}\mu}|0\rangle \otimes |\beta\rangle \otimes |0\rangle + |\psi^\perp\rangle
\end{equation}
where $|\psi^\perp\rangle$ refers to the case that the ancillary qubits do not all give the result `0'. Finally, the post-selection only the results of both ancillary qubits gives `0', the output of the algorithm becomes $|\beta\rangle$ with the success probability of $\frac{1}{2^{m+1}\mu^2}$.

\subsection{Polynomial Root-finding by Eigenvalue Estimation Technique}
 \begin{figure*}[t!]
 	\centering
 	\includegraphics[width=0.9\textwidth]{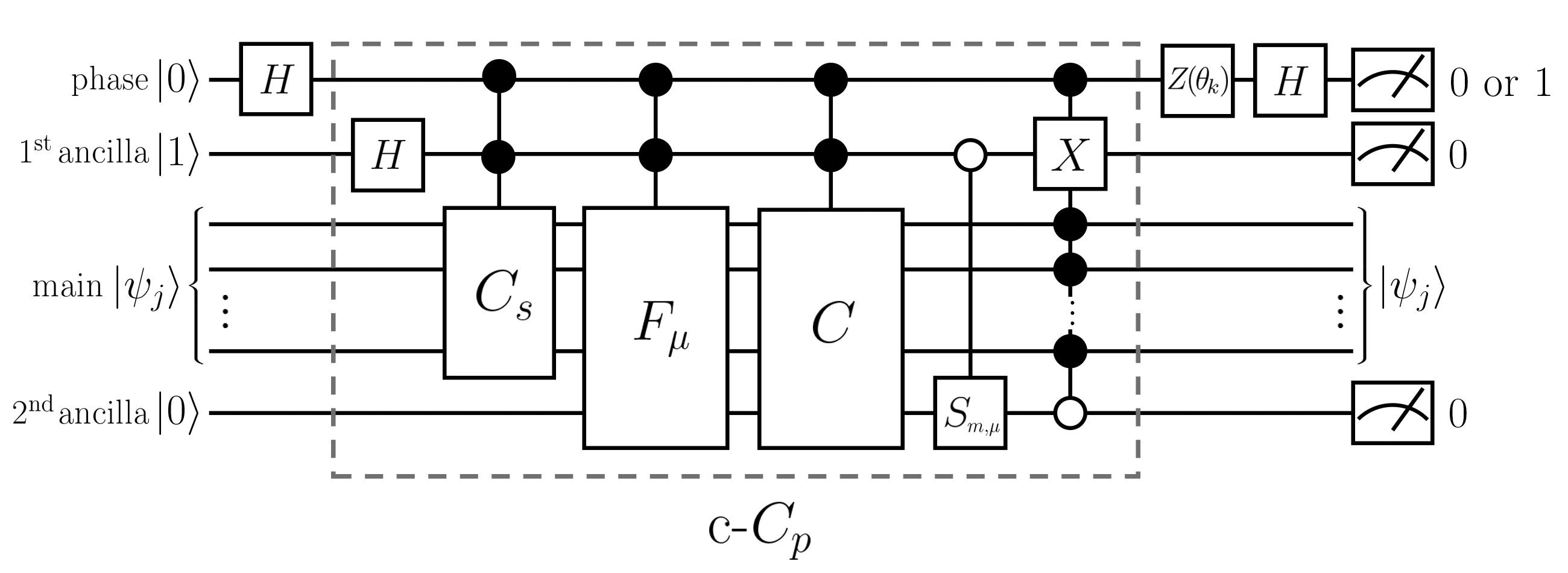}
 	\caption{A circuit scheme for finding polynomial roots.}
 	\label{fig:5}
 \end{figure*}
In order to find roots of the polynomial, we will use the circuit shown in Fig. \ref{fig:5}. A controlled operation of the PRC by the phase qubit is denoted by c-$C_p$ in the figure. To describe the operation, we will firstly assume that the main qubits are initially prepared in an eigenstate $|\psi_j\rangle$ of the companion matrix. An expected state from the operation will be in the form
\begin{equation}
|\beta\rangle = |\lambda_j|\text{e}^{2\pi i\omega_j}|\psi_j\rangle.
\end{equation}

\begin{figure*}[t!]
 	\centering
 	\includegraphics[width=0.9\textwidth]{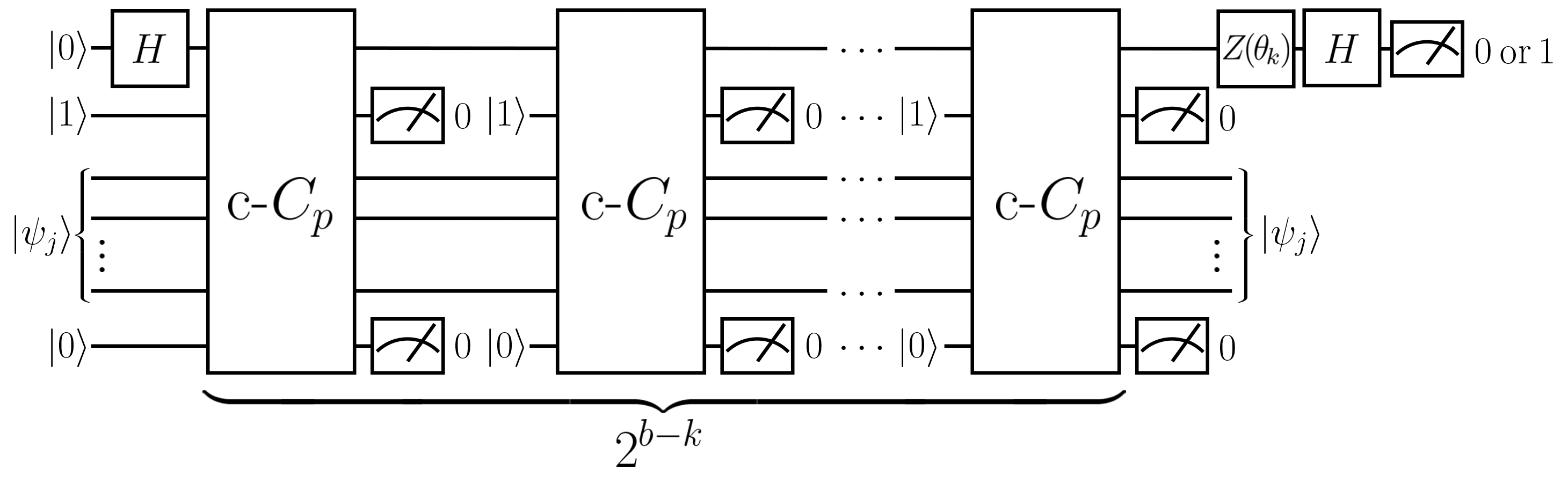}
 	\caption{In order to estimate polynomial roots up to $b$ bit-precision, c-$C_p$ must be operated $2^{b-k}$ times in the $k^{\mathrm{th}}$ iteration.}
 	\label{fig:6}
 \end{figure*}

Here we will assume that $\omega_j$ has a binary expansion in the form $\omega_j = 0.x_1x_2x_3\ldots x_b$, where $b$ is bit-precision. As in IPEA, the c-$C_p$ will be operated $2^{b-k}$ times in the $k^\mathrm{th}$ iteration as illustrated in Fig. \ref{fig:6}. The result of the first iteration is given by
\begin{equation}
	\frac{1}{2}\! \!\left(\!\!\frac{1}{\sqrt{2^{m}}\mu}\!\right)^{2^{b-1}}\!\!\!\!\!\left(\!|0\rangle|0\rangle|\psi_j\rangle|0\rangle\!+\!|\lambda_j|^{2^{b-1}}\!\text{e}^{2\pi i(0.x_b)}|1\rangle|0\rangle|\psi_j\rangle|0\rangle \!\right).
\end{equation}
Similarly, for the $k^{\mathrm{th}}$ iteration, we have
\begin{eqnarray}
	 &&\hspace*{-1cm}\frac{1}{2} \left(\frac{1}{\sqrt{2^{m}}\mu}\right)^{2^{b-k}}\!\Big(\!|0\rangle|0\rangle|\psi_j\rangle|0\rangle \nonumber \\&+&|\lambda_j|^{2^{b-k}}\!\text{e}^{2\pi i(0.x_{b-k+1}x_{b-k+2}\ldots x_b)}|1\rangle|0\rangle|\psi_j\rangle|0\rangle \!\Big).
\end{eqnarray}
After the operation of $Z(\theta_k)$ with $\theta_1=0$ and $\theta_k=2\pi\left(0.0x_{b-k+2}x_{b-k+3}\ldots x_b\right)$ followed by the Hadamard gate, the phase qubit will be in the following state,
\begin{eqnarray}
	&&\hspace*{-1cm}\frac{1}{2\sqrt{2}}\left(\!\frac{1}{\sqrt{2^{m}}\mu}\!\right)^{2^{b-k}}\!\Big(\!(1\!+\!|\lambda_j|^{2^{b-k}}\text{e}^{2\pi i(0.x_{b-k+1})})|0\rangle \nonumber\\ &+&(1\!-\!|\lambda_j|^{2^{b-k}}\text{e}^{2\pi i(0.x_{b-k+1})})|1\rangle \!\Big).
\end{eqnarray}
The value of $x_{b-k+1}$ can be either 0 or 1. Therefore, the probabilities of finding the phase qubit in states $|0\rangle$ or $|1\rangle$ given that the both ancillary qubits give result 0 depend on the value of $x_{b-k+1}$; namely, 
\begin{eqnarray}
	P_0 &=& \frac{1+2\cos(2\pi0.x_{b-k+1})|\lambda_j|^{2^{b-k}}+|\lambda_j|^{2^{b-k+1}}}{8\left(2^m\mu^2\right)^{2^{b-k}}},\\
	P_1 &=& \frac{1-2\cos(2\pi0.x_{b-k+1})|\lambda_j|^{2^{b-k}}+|\lambda_j|^{2^{b-k+1}}}{8\left(2^m\mu^2\right)^{2^{b-k}}}.
\end{eqnarray}
Since $x_{b-k+1}$ can be either 0 or 1, the value of $\cos(2\pi0.x_{b-k+1})$ is either $+1$ or $-1$. In practice, $x_b$ can be obtained by comparing $P_0$ and $P_1$\cite{gr-prl-76-3228}, i.e., $x_b=0$ if and only if $P_0>P_1$; and $x_b=1$ if and only if $P_0<P_1$. In addition, the value of $|\lambda_j|$ from the $k^{\mathrm{th}}$ iteration can be calculated from the equation
\begin{equation}
|\lambda_j|^{2^{b-k+1}}=\left(4\left(2^{m}\mu^2\right)^{2^{b-k}}|P_0+P_1|-1\right).
\end{equation}

It should be emphasized that in later iterations, the parameter $\theta_k$ used in $Z(\theta_k)$ is constructed from $x_{b-k+1}$ from the prior iterations as in IPEA. Finally, an estimate of $\lambda_j$ can now be obtained and the corresponding root of the polynomial can be calculated depending on which mode ($x$-mode or $1/x$-mode) is being used.

However, in general, the eigenstates of the companion matrix are unknown. The following approach is to estimate the greatest eigenvalue $|\lambda_{\mathrm{max}}|$. Suppose that an initial state is prepared in a mixed state, the density operator can be expressed as
\begin{equation}
\rho=\sum_jA_j|\psi_j\rangle\langle\psi_j|,
\end{equation}
where $A_j$ is a probability of preparing the initial state in the eigenstate 
$|\psi_j\rangle$. The operation of c-$C_p$ when the phase qubit is in state $|1\rangle$ transforms the density operator as
\begin{eqnarray}
\rho \mapsto \left(C_p\right)\rho \left(C_p^\dagger\right).
\end{eqnarray}
For the $k^\mathrm{th}$ iteration after which c-$C_p^{2^{b-k}}$ is operated and followed by $Z(\theta_k)$ and the Hadamard gate, the probabilities of finding phase qubit in states $|0\rangle$ or $|1\rangle$ given that both ancillary qubits give result 0 are
\begin{eqnarray}
P_0 &=& \sum_j A_j\left(1\pm2|\lambda_j|^{2^{b-k}}+|\lambda_j|^{2^{b-k+1}}\right)/ \kappa, \label{eq:P0mix} \\
P_1 &=& \sum_j A_j\left(1\mp2|\lambda_j|^{2^{b-k}}+|\lambda_j|^{2^{b-k+1}}\right)/ \kappa. \label{eq:P1mix}
\end{eqnarray}
where $\kappa=8\left(2^m\mu^2\right)^{2^{b-k}}$. In Equations (\ref{eq:P0mix}) and (\ref{eq:P1mix}), the terms with the largest eigenvalue $|\lambda_{\mathrm{max}}|$ will dominate if the number of iterations is large enough. Hence, the probabilities $P_0$ and $P_1$ will be reduced to the following forms:
\begin{eqnarray}
P_0 &\approx& \left[1+A_{\mathrm{max}} \!
\left(|\lambda_{\mathrm{max}}|^{2^{b-k+1}}\!\!\pm2|\lambda_{\mathrm{max}}|^{2^{b-k}}\right)\right]\!\!/\kappa, \\
P_1 &\approx& \left[1+A_{\mathrm{max}} \!
\left(|\lambda_{\mathrm{max}}|^{2^{b-k+1}}\!\!\mp2|\lambda_{\mathrm{max}}|^{2^{b-k}}\right)\right]\!\!/\kappa.
\end{eqnarray}

The value of $x_{b-k+1}$ can be found by comparing $P_0$ and $P_1$. Even without knowing the probability $A_{\mathrm{max}}$, an estimate of $|\lambda_{\mathrm{max}}|$ can be obtained from
\begin{equation}
|\lambda_{\mathrm{max}}|^{2^{b-k}}=2\left(\frac{P_0+P_1-2/\kappa}{|P_0-P_1|}\right).
\end{equation}
After the largest root is found, it can be factorized from the polynomial, and the same technique and procedure can be repeated to calculate the other roots.
 
\section{Discussions}

In order to justify the efficiency of the quantum algorithm for finding roots of a polynomial,
we have to compare it with the classical algorithm for solving the same problem. One of the most efficient classical algorithms for finding roots of a polynomial is created by Pan in 2002\cite{pan-symbcomp-33-701}. We shall compare these two versions of the algorithm based on (i) resources required for calculation and (ii) algorithmic complexity.

We start by comparing the number of bits and qubits required by the computations.
In order to find roots of an $n$\textsuperscript{th} degree polynomial, the quantum algorithm requires $O(\log n)$ qubits. In contrast, Pan's classical root-finding algorithm requires $O(n)$ or $O(n \log n)$ bits. 
This makes it obvious that, for large $n$, the quantum algorithm requires many fewer bits than its classical counterpart.
In this way, the quantum version of algorithm is capable for finding roots of 
a much higher order degree than the classical one.

Next, we will compare their algorithmic complexities. In Pan's algorithm, the number of operations required to find the roots is
\begin{equation} \label{eq:complexC}
	O((n\log^2 n)(\log^2 n + \log b))
\end{equation}
where $b$ is the bit precision of the solutions.
In contrast, since any $m$-qubit unitary gates can be simulated using only single-qubit gates and CNOT gates \cite{ba-prsa-449-679}, 
it is appropriate to compute the complexity of a quantum circuit in terms of the number of single-qubit gates and CNOT gates required to construct the circuit.
From polynomial root-finding circuit, many unitary operations are controlled by several qubits. We will use the following corollary to calculate complexity of these gates (See Corollary 7.12 in Barenco \textit{et al.}\cite{ba-pra-52-3457}).

\textbf{Corollary:} \emph{For any unitary $U$, the corresponding c-$U$ gate controlled by $(m-2)$-qubit can be simulated by $O(m)$ basic operations in $m$-qubit network, where the initial value of one qubit is fixed and incurs no net change.}

In order to apply this corollary with the root-finding circuit, we need one more ancillary qubit and set its initial value to state $|0\rangle$. Note that this ancillary qubit can also be reused to simulate several controlled-unitary operations. In order to compute the overall complexity of the whole circuit, a complexity of each part may be found separately. The cyclic-swap gate can be simulated by $m$ CNOT gates. Number of control qubits of each gate varies from 0 to $m-1$. 
For other CNOT gates with multiple control qubits, its operation can be simulated by $O(i+2)$ operations where $i>1$ is the number of control qubits. 
Therefore the overall complexity of the cyclic-swap gate is $2+\sum_{i=2}^{m-1}i \sim O(m^2)$. 
The formation gate consists of $2^m$ controlled rotation gates. Each gate, which is further controlled by $m$ qubits, can be simulated by $O(m+2)$ basic operations. Hence the total complexity of the formation gate is $O(2^m m)$. The combination gate consists of $m$ Hadamard gates and $m$ NOT gates 
which brings its overall complexity to $O(m)$.
Finally, the complexity of the $(m+1)$-qubit CNOT gate before the application of the last Hadamard gate can be readily computed by the corollary above, which amounts to $O(m)$. 

The total complexity of the complete circuit is the sum of complexity of each part as described above. However, in the complexity calculation, only the greatest term is kept. In the case of large $n$ (where $n$ is the degree of a polynomial), we can clearly see that the dominant term 
comes from the formation gate. Therefore, the complexity of the circuit in terms of the degree of a polynomial is
\begin{equation}
	O(2^m m) \sim O(n\log n),
\end{equation}
for one iteration, or
\begin{equation}
	O(kn\log n),
\end{equation}
for $k$ iterations.
Compared with the classical case for $b$-bit precision, our approach needs $k=2^b$ and the total complexity of the quantum version is
\begin{equation} \label{eq:complexQ2}
	O(2^b n\log n).
\end{equation}
Here, we can validate that the quantum version of the algorithm is less complex than the classical version in the case where the  polynomial has a high degree and a small number bit precision is required. 
On the contrary, the quantum algorithm may be an overkill when finding the roots, with high-bit precision, of a low degree polynomial.

\section{Conclusion}
In summary, we have provided a quantum algorithm for finding roots of the $n^\mathrm{th}$ degree polynomial partially based on Daskin \textit{et al.}'s circuit for finding complex eigenvalues of a general matrix\cite{dask-qip-13-333}. To make a comparison with classical version of the algorithm, 
resources and algorithmic complexities are considered.
The quantum version requires 
fewer number of (quantum) bits than its classical counterpart for a high degree polynomial. In terms of algorithm complexities, the quantum algorithm also 
trumps the classical algorithms for a high degree polynomial, 
requiring low bit-precision solutions. 
The growth in complexity stems from a larger number of iterations 
needed to achieve the desired precision.
Although our result clearly shows that 
finding the roots of polynomials using quantum information scheme is possible, the most important challenge
, however, remains in strengthening the algorithm to overcome the classical algorithm 
both in the utilized resources and the chosen precisions. 
Another challenge 
lies of course in the 
practical issue of a working quantum computer. It is well known that the current quantum computer technology still falls short of the theoretical requirement of the algorithm, especially in terms of the number of entangled qubits and multiple-qubit quantum operations.

\section*{Acknowledgement}
We would like to thank Assist.~Prof.~Dr. Kwan Arayathanitkul for helpful discussions. This work is a collaboration of Collaborative Research Unit on Quantum Information, Mahidol University and Optical and Quantum Communication (OQC) Laboratory, National Electronics and Computer Technology Center (NECTEC). Grant No. 035/2557 from the Development and Promotion of Science and Technology Talents project (DPST) scholarship, research fund for DPST graduate with first placement is acknowledged.

\bibliography{bib15b}
\bibliographystyle{unsrt}

\end{document}